# Spatial geometry of charged rotating and non-rotating rings in rotating and non-rotating frames.

Romannikov A. N.

*Institution "Project Center ITER", Kurchatov sq., 1, 123182, Moscow, Russia*

**Abstract**

Spatial geometry of charged thin rotating and non-rotating rings in a rotating frame is investigated. It is shown, on an example of interaction between a charged probe and two positive charged non-rotating and negative charged rotating rings that the spatial geometry of the rotating ring in the rotating frame has to be different to the spatial geometry of the rotating frame. In the absent of direct relation between the spatial geometry rotating frame and the spatial geometry of the rotating ring in that frame the possibility of a non-flat spatial geometry of rotating electron rings in tokamak plasma is discussed.

1. **Introduction.**

Discussion about spatial geometry of rotating discs or rings in a rotating frame has very long history [see, for example, 1-10]. We will discuss for simplicity rotating and non-rotating thin rings (thin toroids with major radius $R$ and minor radius $r_0$, where $r_0 \ll R$; #1 and #2 rings are non-charged in Introduction) in the article, see Fig.1. It will not be very massive rings with general relativity effects. Starting point of main part discussions is the statement that spatial geometry of rotating rings with radius $R$ in the laboratory flat frame has to be Euclidean. Let us name it "statement 1". It means that the circumference $L$ of the rotating ring is $L = 2\pi R$ in the laboratory non-rotating frame and $L$ has not dependence on the rotating velocity $V$. Main conclusion from this fact is non-Euclidian and the hyperbolic spatial geometry ($L_{rot} > 2\pi R$, where $L_{rot}$ is the circumference of the ring in the rotation frame) of the non-rotating ring #1 in the rotating with velocity $V$ (for $R$) frame [2, 5]. Unfortunately, it is practically impossible to proof one and "statement 1" by experimental observing the real rigid rotating rings because it is impossible to measure very small relativistic effects against the background of the deformation by centrifugal forces for a current accessible $V$ using existed techniques. Probably it is main reason of more than 100 year discussions on this theme with different conclusions. Statements range from the spatial geometry of the non-rotating ring in rotating frame (#1, see Fig.1) being hyperbolic ($L_{rot} > 2\pi R$), to the ring remaining Euclidian

($L_{rot} = 2\pi R$) and to the ring being elliptic ($L_{rot} < 2\pi R$) [6, 9- 13].

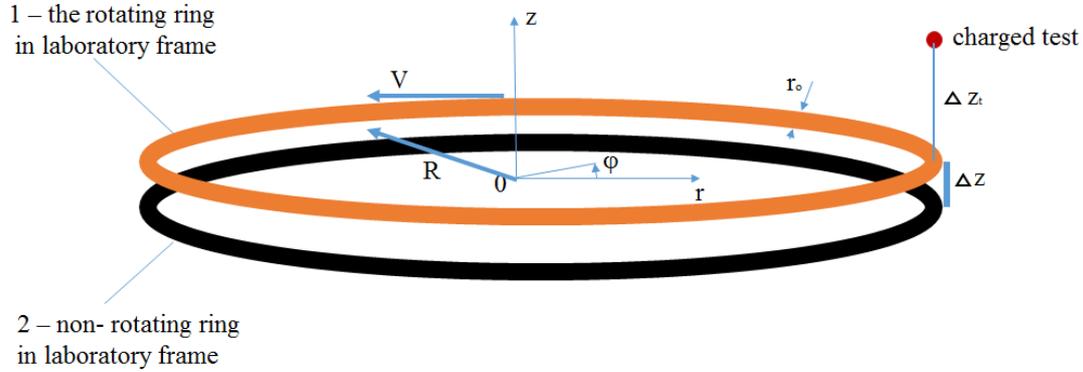

1 – the rotating ring in laboratory frame

2 – non- rotating ring in laboratory frame

charged test

Fig.1. Two rings in the laboratory flat frame.

We will use in our investigation the mainstream very clear discussion of rotating disc behavior presented in [10]. Using cylindrical coordinates ($r$, $\varphi$ and $z$) for laboratory flat frame and time $t$ one can write a squared line element $ds^2$ corresponding of Minkowski metric as:

$$ds^2 = -c^2 dt^2 + dr^2 + r^2 d\varphi^2 + dz^2 \qquad (1)$$

for the space around of the ring; and

$$ds_{\#1}^2 \cong -c^2 dt^2 + R^2 d\varphi^2 \qquad (2)$$

for the points of the ring #1. For simplicity, we separate the metric of the space and the metric of ring points.

Using new rotating coordinates:

$$t_r = t,\ r_r = r,\ \varphi_r = \varphi - \frac{V}{R}\cdot t,\ z_r = z \qquad (3)$$

and for the ring #1: $t_r^{\#1} = t$, $r_r^{\#1} \cong R$, $\varphi_r^{\#1} = \varphi - \frac{V}{R}\cdot t$, $z_r^{\#1} \cong 0$;

one can rewrite (1) for the rotating frame into

$$ds^2 = -c^2 d\tau^2 + dl^2 = -(1 - \frac{V^2 r_r^2}{c^2 R^2}) \cdot \left( cdt_r - \frac{V r_r^2}{cR \cdot (1 - \frac{V^2 r_r^2}{c^2 R^2})} d\varphi_r \right)^2 +$$
$$+ dr_r^2 + r_r^2 \cdot \frac{1}{1 - \frac{V^2 r_r^2}{c^2 R^2}} d\varphi_r^2 + dz_r^2$$

(4)

for the space around of the ring, where $\frac{V^2 r_r^2}{c^2 R^2} < 1$; and

$$ds_{\#1}^2 = -c^2 d\tau_{\#1}^2 + dl_{\#1}^2 \cong$$
$$\cong -(1 - \frac{V^2}{c^2}) \cdot \left( c(dt_r^{\#1}) - \frac{VR}{c \cdot (1 - \frac{V^2}{c^2})} (d\varphi_r^{\#1}) \right)^2 + R^2 \cdot \frac{1}{1 - \frac{V^2}{c^2}} (d\varphi_r^{\#1})^2$$

(5)

for the points of the ring #1.

We will study mainly the spatial geometry squared line element $dl^2$. Main conclusion from (2) and (5) is starting from flat spatial geometry ($L = 2\pi R$) the ring #1 to change the spatial geometry in the rotating frame to hyperbolic ($L_{rot} > 2\pi R$). Physically this result is very clear. Let us divide the circumference of the #1 ring in the rotating frame on N segments. According of (5) one can get the length of a segment $\Delta L_r = \frac{L_{rot}}{N} = \frac{2\pi R}{N \cdot \sqrt{(1 - \frac{V^2}{c^2})}} = \frac{2\pi R \cdot \gamma}{N}$. We can proposal that small segment belongs to a local inertial frame. This frame and the segment move with velocity $V$ to the inertial non-rotating laboratory frame short time. According of Lorentz contraction the length of the segment in non-rotating frame will be $\Delta L = \Delta L_r \cdot \frac{1}{\gamma} = \frac{2\pi R \cdot \gamma}{N \cdot \gamma} = \frac{2\pi R}{N}$ and the circumference of the rotating ring #1 in the non-rotating laboratory frame will be $L = 2\pi R$.

One can use another way for creation of the circumference $L_{rot} = 2\pi R \cdot \gamma$. The usually we accelerate initially non-rotating ring from zero velocity to $V(t)$. One can say that the acceleration program of ring points in the rotating frame with $V(t)$ on the ring fixes the segment length by $\Delta L_r(t) = \frac{2\pi R \cdot \gamma(t)}{N} \cong \gamma(t) \cdot R d\varphi_r^{\#1}$ during acceleration. It is in the frame of Feynberg' ideas presented in [14]. We have $L_{rot} = 2\pi R \cdot \gamma$ at the end. It is like Bell's type acceleration program [10, 15, 16] but for the rotating frame with $V(t)$ on the ring #1. We name that procedure "Bell's type acceleration program" - BTAP for future. This program is equivalent to the Bell acceleration program with fixation of the #1 ring circumference as $L = \Delta L \cdot N = 2\pi R$ in the laboratory frame.

The analysis of the ring #2 spatial geometry in the rotating frame is more complex. The author [10] shown for the ring #2 that

$$d\varphi = \gamma^2 d\varphi_r \qquad (6)$$

assuming of "Einstein synchronization". If $\varphi$ runs from 0 to $2\pi$, $\varphi_r$ will only cover an interval of length $\dfrac{2\pi}{\gamma^2}$. This is smaller than $2\pi$ and the non-rotating ring #2 does not fully cover the rotating ring #1 in the rotating frame. Because there no discontinuity, the author [10] proposed that the continuation was carried out another copy of the ring #2 but later. According of the author [10] "Einstein synchronization" was main reason of particular behavior of the non-rotating ring #2 in the rotating frame. It was shown in [10] that, using another type synchronization - "central synchronization", the spatial geometry of the ring #2 in the rotating frame becomes equivalent to the spatial geometry of the ring #1 in the rotating frame. It means that the spatial geometry of the ring #1 and #2 in the rotating frame is hyperbolic and

$$L_{rot}^{\#1} \equiv L_{rot}^{\#2} = 2\pi R \cdot \gamma > 2\pi R \qquad (7).$$

2. **Spatial geometry of charged rotating and non-rotating rings in the rotating frame**

Let us investigate the case with charged #1 and #2 rings. Let the charge of the #1 ring is $Q_1 = -|e| \cdot n \cdot V_r$ and the charge of the #2 ring is $Q_2 = |e| \cdot n \cdot V_r$, where $e$ is the charge of electron, $n$ is the homogenous density of charge in the laboratory frame and $V_r$ is the full volume of the each ring. Charged probe is positive $+q$. Let $\Delta z \ll \Delta z_t \ll R$ and we can neglect the dipole part of electric and magnetic fields in the point of the charged probe and we do not pay attention to some additional mechanical forces that necessary for equilibrium. It is enough difficult to carry out that experiment in reality. Nevertheless, it will be shown later the example with two charged rings that occupies practically the same space. We will neglect terms of

expansion more that $\frac{V^2}{c^2}$ in our calculations below. It is clear that the electric field of the #1 ring ( $E_1^0(\Delta z_t)$ ) in the point of the charged probe is practically equal of the electric field of the #2 ring ( $E_2^0(\Delta z_t)$ ), so

$$E^0(\Delta z_t) = E_1^0(\Delta z_t) + E_2^0(\Delta z_t) \cong -\frac{4\pi|e|\cdot n\cdot}{\Delta z_t}\cdot\int_0^{r_0}\xi\cdot d\xi + \frac{4\pi|e|\cdot n\cdot}{\Delta z_t}\cdot\int_0^{r_0}\xi\cdot d\xi =$$

$$= -\frac{2\pi|e|\cdot n\cdot r_0^2}{\Delta z_t} + \frac{2\pi|e|\cdot n\cdot r_0^2}{\Delta z_t} = 0$$

Only the rotating ring #1 creates the magnetic field $H^0(\Delta z_t) = H_1^0(\Delta z_t) \cong -\frac{4\pi|e|\cdot n\cdot V}{\Delta z_t\cdot c}\cdot\int_0^{r_0}\xi\cdot d\xi = \frac{2\pi|e|\cdot n\cdot V\cdot r_0^2}{\Delta z_t\cdot c}$.

The sum electromagnetic force in the point of the charged probe is zero in our proposals. Let us study electromagnetic forces in the rotating frame. Here only the rotating ring #2 creates the magnetic field $H^r(\Delta z_t) = H_2^r(\Delta z_t)$. The sum electric field in this frame is $E^r(\Delta z_t) = E_1^r(\Delta z_t) + E_2^r(\Delta z_t) \cong -\frac{2\pi|e|\cdot n\cdot r_0^2}{\Delta z_t\cdot\gamma} + \frac{2\pi|e|\cdot n\cdot r_0^2}{\Delta z_t\cdot\gamma} = 0$. We have decreasing of charge densities on the factor $\frac{1}{\gamma}$ for #1 and #2 rings according of eq. 7. However, the charge probe rotates in the rotating frame and we have a non-compensated force for one on $z$ direction proportional to $-q\cdot\frac{V}{c}H^r(\Delta z_t) \cong -q\cdot\frac{V}{c}\frac{2\pi|e|\cdot n\cdot V\cdot r_0^2}{\Delta z_t\cdot c\cdot\gamma}$. The charged probe is obliged to move initially on $z$ direction in the rotating frame and we will see this movement in the laboratory frame too. We have the evident contradiction.

For compensation of this force, it is necessary to have some non-compensating
$$E^r(\Delta z_t) = E_1^r(\Delta z_t) + E_2^r(\Delta z_t) \neq 0 \tag{8}$$

According of eq. 7 it is not possible.

This contradiction could be eliminated using eq. 6, but another manner, at BTAP for case with the #2 ring.

Physically in the case with the #2 ring we create the new rotation frame with $\gamma(t)$ on the #2 ring starting from the laboratory flat frame with $\gamma(t_0) = 1$. The final state is the rotating frame with

$\gamma = \dfrac{1}{\sqrt{1-\dfrac{V^2}{c^2}}}$ on the #2 ring. During this acceleration we use BTAP with fixation points of the

#2 ring as $\Delta L_r(t) \cong \dfrac{1}{\gamma(t)^2} \cdot Rd\varphi_r^{\#2}$, according of eq. 6. It can change the squared line element of the #2 ring points as:

$$ds_{\#2}^2 = -c^2 d\tau_{\#2}^2 + dl_{\#2}^2 \cong$$
$$\cong -(1-\dfrac{V^2}{c^2}) \cdot \left( c \cdot (dt_r^{\#2}) - \dfrac{V}{c} R \cdot (d\varphi_r^{\#2}) \right)^2 + R^2 \cdot (1-\dfrac{V^2}{c^2})(d\varphi_r^{\#2})^2 \qquad (9).$$

It means that $L_{rot}^{\#1} = 2\pi R \cdot \gamma \neq L_{rot}^{\#2} = \dfrac{2\pi R}{\gamma}$.

Of course derivation of the eq. (9) is more physical than strong mathematical, but we having no contradiction for behavior of charged rings with the charged probe in the rotation frame. Indeed

$$E^r(\Delta z_t) = E_1^r(\Delta z_t) + E_2^r(\Delta z_t) \cong -\dfrac{2\pi|e|\cdot n \cdot r_0^2}{\Delta z_t \cdot \gamma} + \dfrac{2\pi|e|\cdot n \cdot \gamma \cdot r_0^2}{\Delta z_t} = \gamma \dfrac{2\pi|e|\cdot n \cdot r_0^2}{\Delta z_t} \dfrac{V^2}{c^2} \qquad (10).$$

The magnetic field in that point is $H^r(\Delta z_t) \cong \dfrac{2\pi|e|\cdot n \cdot \gamma \cdot V \cdot r_0^2}{\Delta z_t \cdot c}$. The electromagnetic force on the charge probe in the rotating frame is $q \cdot E^r(\Delta z_t) - q \cdot \dfrac{V}{c} H^r(\Delta z_t) = 0$ in frames of our approximations.

We can do the main conclusion now. The charge density of the #2 ring in the rotating frame has to change as $n \Rightarrow \gamma \cdot n$. The simple way for one is the statement: "The special geometry of the charged rotating #2 ring in the rotating frame, eq. 9, has to be different to the special geometry of the charged non-rotating #1 ring in the rotating frame, eq. 5, and to the special geometry of the rotating space around rings, eq. 4".

## 3. Two charged rings in tokamak plasma

An interesting question is appeared with the conclusion of the previous paragraph. If the relation between the spatial geometry rotating frame and the spatial geometry of the rotating ring in that frame is particular, can be created "artifact" in the laboratory flat frame with the different spatial geometry to spatial geometry around that "artifact"? In other words in the frame of our discussion, is "statement 1" the law of nature or is there exception to that law? Below it will be shown an example of possible exception to "statement 1".

Recently, in [17, 18, 19] the example two charged rings filling the same space in a tokamak plasma has been presented. Simple sketch of tokamak is presented on Fig.2. The minor radiuses a tokamak magnetic surface $r$ and the radius of the metallic tokamak chamber $a$ was assumed to be much less than the major radius $R$, where $\dfrac{r}{R} < \dfrac{a}{R} \ll 1$.

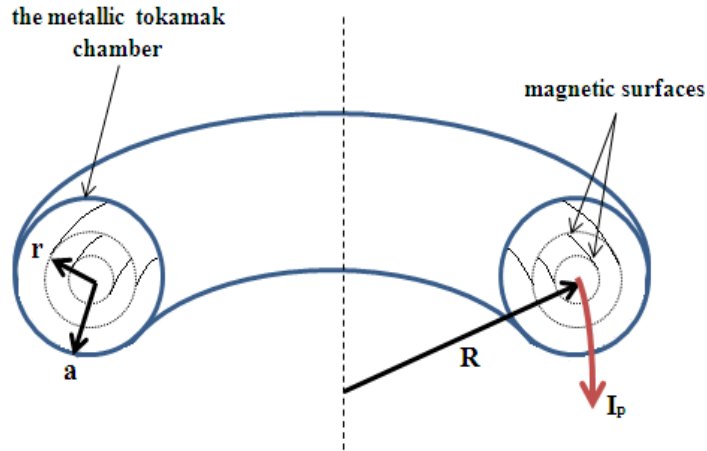

**Fig.2.** A sketch of tokamak.

It is assumed that at the moment of plasma creation (with no current) from neutral gas (hydrogen or deuterium), the electron density $n_e^0(r)$ and the ion density $n_i^0(r)$ are equal, and the difference between the total number of electrons and total number of ions does not vary during a discharge and is equal to 0 in the tokamak chamber. We can use the last assumption because neutral gas is injected into the tokamak chamber, and neutral gas is pumped from the tokamak. Electrons and ions can move and can be redistributed in the minor radius direction of tokamak plasma after the occurrence of a current. But $n_e^0(r) \cong n_i^0(r)$ during a discharge. For our purpose we can consider of tokamak plasma as nested magnetic surfaces with plasma distributed in thin, hollow rings. A vortex electric field creates plasma current $I(r)_p$ with the toroidal electron plasma current velocity $V_e(r)$. The equilibrium toroidal rotation velocity of ions $V_i(r)$, as a general rule, is much less than $V_e(r)$ and we will propose that $V_i(r) \approx 0$. We can say that tokamak plasma during a discharge is the nested set of thin hollow rotating electron rings and non-rotating ion rings. The current electron velocity $V_e(r)$ can reach several hundred km/s. The relativistic effect $\frac{V^2}{c^2}$ of rotating electron rings is in $\sim 10^5$ more than one in the case of rigid rotating rings or discs with current accessible velocities. Let us consider the creation of some tokamak nested electron ring with the minor radius $r_0$ in the rotating frame with $V_e(r_0,t)$ from $V_e(r_0,0) = 0$ to $V_e(r_0,t_{saturation}) = V_e(r_0)$. The ion ring has $V_i(r_0,t) \cong 0$ in the laboratory flat frame and obtains the velocity $V_i^r(r_0,t) \cong V_e(r_0,t)$ in the rotating frame. At acceleration program of the electron ring in the rotating frame with $V_e(r_0,t)$ participates two forces. The first it is the vortex electric field force, the second it is the friction force between the electron ring and the ion ring. According of BTAP we could say that the vortex electric field force tries to fix the segment length $\Delta L_r^e(t) = \frac{2\pi R \cdot \gamma_e(t)}{N} \cong \gamma_e(t) \cdot R d\varphi_r^e$ during acceleration. We use for $\gamma_e(t)$ the velocity $V = V_e(r_0,t)$. But the friction force of the electron ring with the rotating ion ring in the rotating frame tries to decrease the electron segment length to $\Delta L_r^e(t) = \frac{2\pi R}{N \cdot \gamma_e(t)} \cong \frac{1}{\gamma_e(t)} R d\varphi_r^e$. We can expect some equilibrium condition for the electron segment length $\Delta L_r^e(t) \sim \frac{2\pi R}{N} \cong R d\varphi_r^e$ during acceleration as a result of BTAP. It means that the special geometry of the non-rotating electron ring in the rotating frame can be modified to

$$ds_e^2 = -c^2 d\tau_e^2 + dl_e^2 \cong$$
$$\cong -(1 - \frac{V_e^2(r_0,t)}{c^2}) \cdot \left( c \cdot (dt_r^e) - \sqrt{1 - \frac{V_e^2(r_0,t)}{c^2}} \frac{V(r_0,t)}{c} R \cdot (d\varphi_r^e) \right)^2 + R^2 \cdot (d\varphi_r^e)^2$$

and is different to the special geometry of the rotating frame, eq. (4) with $V = V_e(r_0,t)$.

As a result of that type BTAP for the electron ring the special geometry of that ring in the laboratory flat frame will be modified, too and $dl_{e,laboratory}^2 \cong (1 - \frac{V_e^2(r_0,t)}{c^2}) \cdot R^2 \cdot (d\varphi^e)^2$ with

$$L_{laboratory}^i = 2\pi R \neq L_{laboratory}^e = \frac{2\pi R}{\gamma_e} \tag{11}.$$

The possible existing of electron rings with elliptic special geometry in the laboratory frame has some important consequences for tokamak plasma [18, 19].

The first, for tokamak plasma contained in a metallic chamber, the effect of eq. (11) can modify the electric potential of the chamber as

$$\Delta\varphi \cong \frac{\int_{V_p}(-\frac{j^2(r)}{2\cdot c^2 \cdot |e|n_e(r)})\cdot dV_p}{C} \qquad (12)$$

with respect to the ground. The electric potential of the chamber is proportional in this case to the volume of plasma $V_p$, the averaged value of $-\frac{j(r)^2}{2\cdot c^2 \cdot |e|n_e}$, where $j(r)$ is the density of the plasma current, the electric capacitance of a closed metallic tokamak chamber C, and relates to chamber RC time. The measurements of that electric potential have been presented in [17, 19]. The experimental results satisfactory coincided with eq. (12). The $\Delta\varphi$ was ~ 50 V for experiments in small T-11M tokamak. The $\Delta\varphi$ can reach value ~ 1 kV for big tokamak like JET.

The second, radial electric field $E_r(r)$ plays an important role in various modes of improved plasma confinement in a tokamak [20]. Some of those modes will be used in the thermonuclear reactor ITER [21], which is currently under construction. Eq. (11) plays important function in creation of $E_r(r)$ in tokamak. Specific evidences of that influence at tokamak plasma are presented in the articles [18, 19], too. The relativistic theory [17, 18, 19, 22] of radial electric field formation, based on eq. (11), can explain quantitatively and more often qualitatively, many experimental tokamak results.

## 4. Conclusion

Investigation of rotating charged rings shows:

1. On the example of interaction between the charged probe and two positive charged non-rotating and negative charged rotating rings, the spatial geometry of the rotating ring #2 in the rotating frame has to be different to the spatial geometry of the rotating frame.
2. The circumference the ring #1 differs to the circumference of the ring #2 in the rotating frame as: $L_{rot}^{\#1} = 2\pi R\cdot\gamma \neq L_{rot}^{\#2} = \frac{2\pi R}{\gamma}$. It give us the electric field in the rotating frame, eq. (10), which creates the necessary compensation force for equilibrium of the charged probe in the rotation frame.

3. There is the possible exception to "statement 1". It is rotating electron current rings in tokamak plasma. Rotating electron rings and non-rotating ion rings situated the same space in tokamak plasma change the radial electric field in regions close to rings. Experimental evidences [18, 19, 20] in tokamak plasmas confirm with that exception.